\documentstyle[12pt,a4,psfig,axodraw]{article}
\begin{document}
\baselineskip 18pt
\def\today{\ifcase\month\or
 January\or February\or March\or April\or May\or June\or
 July\or August\or September\or October\or November\or December\fi
 \space\number\day, \number\year}
\def\thebibliography#1{\section*{References\markboth
 {References}{References}}\list
 {[\arabic{enumi}]}{\settowidth\labelwidth{[#1]}
 \leftmargin\labelwidth
 \advance\leftmargin\labelsep
 \usecounter{enumi}}
 \def\newblock{\hskip .11em plus .33em minus .07em}
 \sloppy
 \sfcode`\.=1000\relax}
\let\endthebibliography=\endlist
\def\MPLA#1#2#3{Mod. Phys. Lett. {\bf A#1} (#2) #3}
\def\PRD#1#2#3{Phys. Rev. {\bf D#1} (#2) #3}
\def\NPB#1#2#3{Nucl. Phys. {\bf B#1} (#2) #3}
\def\PTP#1#2#3{Prog. Theor. Phys. {\bf #1} (#2) #3}
\def\ZPC#1#2#3{Z. Phys. {\bf C#1} (#2) #3}
\def\EPJC#1#2#3{Eur. Phys. J. {\bf C#1} (#2) #3}
\def\PLB#1#2#3{Phys. Lett. {\bf B#1} (#2) #3}
\def\PRL#1#2#3{Phys. Rev. Lett. {\bf #1} (#2) #3}
\def\PRep#1#2#3{Phys. Rep. {\bf #1} (#2) #3}
\def\RMP#1#2#3{Rev. Mod. Phys. {\bf #1} (#2) #3}
\def\gsim{~{\rlap{\lower 3.5pt\hbox{$\mathchar\sim$}}\raise 1pt\hbox{$>$}}\,}
\def\lsim{~{\rlap{\lower 3.5pt\hbox{$\mathchar\sim$}}\raise 1pt\hbox{$<$}}\,}
\def\ov{\overline}
\def\wt{\widetilde}
\def\r2{\sqrt 2}
\def\half{\frac{1}{2}}
\def\bsg{b\rightarrow s\gamma}
\def\bsgg{b\rightarrow s\gamma g}
\def\BSgamma{B\rightarrow X_s\gamma}
\def\mqu#1{m_{u#1}}
\def\mqd#1{m_{d#1}}
\def\VT{V_{32}^*V_{33}}
\def\VU{V_{42}^*V_{43}}
\def\ZBS{(V^\dagger V)_{23}}
\newcommand{\beq}{\begin{equation}}
\newcommand{\eeq}{\end{equation}}
\newcommand{\bea}{\begin{eqnarray}}
\newcommand{\eea}{\end{eqnarray}}
\begin{titlepage}
\begin{flushright}
\begin{tabular}{l}
{OCHA-PP-155} \\
{KEK-TH-693}\\
\end{tabular}
\end{flushright}
\vskip 0.5 true cm 
\begin{center}
{\large {\bf Contributions of vector-like quarks  \\
to radiative $B$ meson decay}}  
\vskip 2.0 true cm
\renewcommand{\thefootnote}
{\fnsymbol{footnote}}
Mayumi Aoki$^1$, Eri Asakawa$^2$, Makiko Nagashima$^2$, 
Noriyuki Oshimo$^3$, \\ 
and Akio Sugamoto$^{2,3}$ 
\\
\vskip 0.5 true cm 
{\it $^1$Theory Group, KEK, Tsukuba, Ibaraki 305-0801, Japan}  \\
{\it $^2$Graduate School of Humanities and Sciences}  \\
{\it Ochanomizu University, Otsuka 2-1-1, 
Bunkyo-ku, Tokyo 112-8610, Japan}  \\
{\it $^3$Department of Physics}  \\
{\it Ochanomizu University, Otsuka 2-1-1, 
Bunkyo-ku, Tokyo 112-8610, Japan}  \\
\end{center}

\vskip 3.0 true cm

\centerline{\bf Abstract}
\medskip
     We study the decay $\BSgamma$ in 
a minimal extension of the standard model with extra   
up- and down-type quarks whose left- and right-handed 
components are both SU(2) singlets.   
Constraints on the extended Cabibbo-Kobayashi-Maskawa matrix 
are obtained from the experimental results for the branching ratio.   
Even if the extra quarks are too heavy to be detected in 
near-future colliders, the branching ratio could have a value which is 
non-trivially different from the prediction of the standard model.  

\vspace{3mm}
\noindent
PACS number(s): 12.15.Ff, 12.60.-i, 13.25.Hw, 13.40.Hq
\medskip

\end{titlepage}

\newpage 

     The inclusive decay $\BSgamma$ is 
well described by the free quark decays $\bsg$ and $\bsgg$, 
owing to a large mass of the $b$ quark.  
Since these decays are generated at the one-loop level 
of the electroweak interactions, 
the radiative $B$-meson decay is sensitive to new physics beyond  
the Standard Model (SM) \cite{nir}, 
such as the supersymmetric model \cite{oshimo}.  
Its branching ratio could deviate from the prediction of the SM.  
Or some constraints could be imposed on new physics.  
Experimentally, the branching ratio has been measured by 
CLEO \cite{cleo} and ALEPH \cite{aleph} as  
\begin{eqnarray}
{\rm Br}(\BSgamma) &=& 
                      (3.15 \pm 0.35 \pm 0.32 \pm 0.26) \times 10^{-4},  
\label{cleo}  \\
      &=& (3.11 \pm 0.80 \pm 0.72) \times 10^{-4}.    
\label{aleph}
\end{eqnarray}
These results are consistent with the SM prediction 
${\rm Br}(\BSgamma) = (3.29\pm 0.33)\times 10^{-4}$ \cite{kagan}, 
though still show room for the contribution of new physics.   

     The SM is minimally extended by incorporating extra colored 
fermions whose left-handed components, 
as well as right-handed ones, are singlets under the SU(2) gauge 
transformation, with electric charges being 2/3 and/or $-1/3$.  
In this vector-like quark model (VQM), 
many features of the SM are not significantly modified.  
However, the interactions of the quarks with the $W$ or $Z$ boson 
are qualitatively different from those in the SM.   
The Cabibbo-Kobayashi-Maskawa (CKM) matrix for the charged 
current is extended and not unitary.  
The neutral current involves 
interactions between the quarks with different flavors.  
In addition, the neutral Higgs boson also mediates 
flavor-changing interactions at the tree level.  
The VQM could thus give sizable new contributions to processes of 
flavor-changing neutral current (FCNC) \cite{branco,morozumi} 
and of $CP$ violation \cite{asakawa,uesugi}.  

     In this paper we study the radiative $B$-meson decay 
within the framework of the VQM containing  
one up-type and one down-type extra quarks.  
The decay receives contributions from the interactions 
mediated by the $W$, $Z$, and Higgs bosons.  
The effects by the $Z$ and Higgs bosons 
have already been studied and found to be small \cite{morozumi}.  
Our study is concentrated on the other effects coming from 
the $W$-mediated interactions.   
These interactions give contributions differently from the SM 
at the electroweak energy scale, 
since an extra up-type quark is involved and the 
CKM matrix is not the same as that of the SM.  
It will be shown that the decay width can be much different 
from the SM prediction, even if the extra quark is rather heavy.  
The experimental results for the decay rate thus impose 
non-trivial constraints on the extended CKM matrix.  

     We assume that there exist two extra Dirac fermions 
whose transformation properties are given by $(3,1,2/3)$ and $(3,1,-1/3)$ 
for the SU(3)$\times$SU(2)$\times$U(1) gauge symmetry.  
The mass terms of the quarks are then expressed by 4$\times$4 matrices.  
These mass matrices, which are denoted 
by $M^u$ and $M^d$ respectively for up- and down-type quarks, 
are diagonalized by unitary matrices 
$A^u_L$, $A^u_R$, $A^d_L$, and $A^d_R$ as 
\begin{eqnarray}
      A_L^{u\dagger} M^uA^u_R &=& {\rm diag}(\mqu1,\mqu2,\mqu3,\mqu4),   \\
      A_L^{d\dagger} M^dA^d_R &=& {\rm diag}(\mqd1,\mqd2,\mqd3,\mqd4).    
\end{eqnarray}
The mass eigenstates are expressed by $u^a$ and $d^a$ ($a=1-4$),   
$a$ being the generation index, which are also called as 
$(u,c,t,U)$ and $(d,s,b,D)$.    

     The interaction Lagrangian for the quarks with the $W$ and 
Goldstone bosons is given by 
\begin{eqnarray}
 {\cal L} &=& \frac{g}{\r2}\sum_{a,b=1}^4\overline{u^a}V_{ab}\gamma^\mu               
                   \frac{1-\gamma_5}{2}d^bW_\mu^\dagger   \nonumber  \\ 
 & & +\frac{g}{\r2}\sum_{a,b=1}^4\overline{u^a}V_{ab}\left\{\frac{m_{ua}}{M_W}            
                   \left(\frac{1-\gamma_5}{2}\right)
              - \frac{m_{db}}{M_W}\left(\frac{1+\gamma_5}{2}\right)\right\}
                d^bG^\dagger  \nonumber \\
         & & +{\rm h.c.}.    
\label{Wlagrangian}
\end{eqnarray}
Here the $4\times 4$ matrix $V$ stands for an extended 
Cabibbo-Kobayashi-Maskawa matrix, which is defined by 
\begin{equation}
      V_{ab} = \sum_{i=1}^3(A_L^{u\dagger})_{ai}(A_L^d)_{ib}.  
\end{equation}
It should be noted that $V$ is not unitary:   
\beq
  (V^\dagger V)_{ab}=\delta_{ab}-A^{d*}_{L4a}A^d_{L4b}.  
\eeq
The interaction Lagrangian for the down-type quarks 
with the $Z$, Higgs, and Goldstone bosons is given by  
\begin{eqnarray}
{\cal L} &=& -\frac{g}{\cos\theta_W}\sum_{a,b=1}^4
 \overline{d^a}\gamma^\mu
            \left\{-\half\left(V^\dagger V\right)_{ab}\frac{1-\gamma_5}{2}
    +\frac{1}{3}\sin^2\theta_W\delta_{ab}\right\}d^bZ_\mu   
\nonumber  \\
   & & -\frac{g}{2}\sum_{a,b=1}^4\overline{d^a}\left(V^\dagger V\right)_{ab}
\left\{\frac{m_{da}}{M_W}\left(\frac{1-\gamma_5}{2}\right)
    +\frac{m_{db}}{M_W}\left(\frac{1+\gamma_5}{2}\right)\right\}d^bH^0   
\nonumber  \\
     & & +i\frac{g}{2}\sum_{a,b=1}^4\overline{d^a}\left(V^\dagger V\right)_{ab}
\left\{\frac{m_{da}}{M_W}\left(\frac{1-\gamma_5}{2}\right)
  -\frac{m_{db}}{M_W}\left(\frac{1+\gamma_5}{2}\right)\right\}d^bG^0.  
\label{Zlagrangian} 
\end{eqnarray}
Since $V$ is not a unitary matrix, there appear interactions of 
FCNC at the tree level.     
The Lagrangians in Eqs. (\ref{Wlagrangian}) and (\ref{Zlagrangian}) 
contain new sources of $CP$ violation \cite{asakawa}.    

\begin{figure}
\begin{center}
\begin{picture}(250,200)(0,0)
\Line(-30,100)(70,100)
\PhotonArc(20,100)(30,0,180){4}{8.5}
\Text(-35,100)[]{$b$}
\Text(75,100)[]{$s$}
\Text(20,109)[]{$u\; c\;t \; U$}
\Text(5,140)[]{$W$}
%
\Line(100,100)(200,100)
\PhotonArc(150,100)(30,0,180){4}{8.5}
\Text(95,100)[]{$b$}
\Text(205,100)[]{$s$}
\Text(150,109)[]{$d\; s\; b\; D$}
\Text(135,140)[]{$Z$}
%
\Line(230,100)(330,100)
\DashCArc(280,100)(30,0,180){4}
\Text(225,100)[]{$b$}
\Text(335,100)[]{$s$}
\Text(280,109)[]{$d\; s\; b\; D$}
\Text(265,140)[]{$H^0$}
\end{picture}
\end{center}
\caption{The diagrams which give contributions to $C_7$ and $C_8$.  
The photon or gluon line should be attached appropriately.}
\label{figloop}
\end{figure}
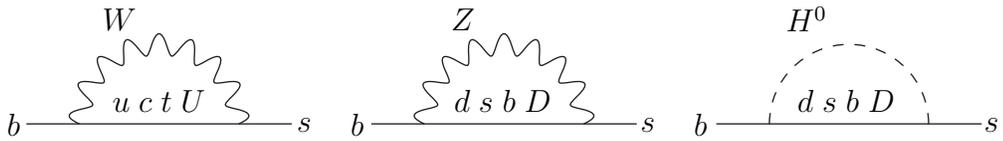

     The decay $\BSgamma$ is approximated by the radiative 
$b$-quark decays, which are mediated by the $W$, $Z$, and Higgs bosons.  
The relevant effective Hamiltonian with five quarks 
is then written as 
\beq
{\cal H}_{eff} = -\frac{4G_F}{\r2} \left[
	\sum_{j=1}^6\left\{C_j(\mu) O_j(\mu)+\wt C_j(\mu)\wt O_j(\mu)\right\} 
+\sum_{j=7}^8C_j(\mu) O_j(\mu)\right],  
\eeq
where $O_j$, $\wt O_j$ represent operators for 
the $\Delta B = 1$ transition, with $C_j$, $\wt C_j$ being 
their Wilson coefficients.
The evaluated energy scale is denoted by $\mu$. 
The four-quark operators induced by the gauge boson 
interactions are denoted by $O_j (j=1-6)$ \cite{buras}.  
The Higgs boson interactions induce new four-quark operators, which are 
denoted by $\wt O_j (j=1-6)$.  
The dipole operators for $\bsg$ and $b\rightarrow sg$ are denoted 
by $O_7$ and $O_8$, respectively, which are generated by the one-loop 
diagrams shown in Fig. \ref{figloop}.  
Hereafter, we only take the $W$ boson interactions into consideration, 
since the contributions coming from the $Z$ and Higgs boson interactions 
are known to be much smaller than the SM contribution.  
        
     At the leading order (LO), the Wilson coefficients 
$C_2$, $C_7$, and $C_8$ have non-vanishing values at $\mu=M_W$, 
which are given by 
\bea
C_2(M_W)&=& V^{\ast}_{32}V_{33}+V^{\ast}_{42}V_{43}-\ZBS, \\
C_7(M_W)&=&\frac{23}{36}\ZBS
            -\sum\limits_{a=3}^4V^{\ast}_{a2}V_{a3}
                \frac{3}{2}r_a\left\{\frac{2}{3}I_1(r_a)+J_1(r_a)\right\}, \\
C_8(M_W)&=&\frac{1}{3}\ZBS
            -\sum\limits_{a=3}^4V^{\ast}_{a2}V_{a3}
                          \frac{3}{2}r_aI_1(r_a), \\
 r_a&=&\frac{m_{ua}^2}{M_W^2}.  \nonumber 
\eea
The functions $I_1(r)$ and $J_1(r)$ are defined as \cite{aoki}
\bea
I_1(r) &=& \frac{1}{12(1-r)^4}(2+3r-6r^2+r^3+6r\ln r), \\
J_1(r) &=& \frac{1}{12(1-r)^4}(1-6r+3r^2+2r^3-6r^2\ln r).  
\eea
The non-unitarity of the CKM matrix $V$ 
yields the terms proportional to $\ZBS$ for $C_2$, $C_7$, and $C_8$.   
The Wilson coefficients at $\mu = m_b$ are obtained by solving 
the renormalization group equations.   
Using the LO anomalous dimension matrix, the coefficients are given by 
\bea
C_2(m_b) &=& \frac{1}{2} (\eta^{-\frac{12}{23}} + \eta^{\frac{6}{23}} )
C_2(M_W),  \\
C_7(m_b) &=& \eta^{\frac{16}{23}} C_7(M_W) + 
 \frac{8}{3} (\eta^{\frac{14}{23}} 
- \eta^{\frac{16}{23}}) C_8(M_W) + \sum_{i=1}^8 h_i \eta^{a_i}C_2(M_W), 
\label{c7} \\
C_8(m_b) &=& \eta^{\frac{14}{23}} C_8(M_W) + 
\sum_{i=1}^8 \bar{h}_i \eta^{a_i}C_2(M_W),  
\label{c8}
\eea
with $\eta= \alpha_s(M_W)/\alpha_s(m_b)$ which is set for $\eta = 0.56$ 
in the following numerical study.
The constants $h_i, \bar{h}_i$, and $a_i$ are listed  
in Table \ref{tabcoef} \cite{buras2}.  
The branching ratio for $\BSgamma$ is obtained by normalizing 
the decay width to that of the semileptonic decay 
$B \to X_ce\overline\nu$,   
leading at the LO to 
\bea
{\rm Br}(\BSgamma)&=&\frac{6\alpha_{\rm EM}}{\pi f(z)|V_{23}|^2}
\left|C_7(m_b)\right|^2{\rm Br}(B \to X_ce\overline\nu),   
\label{LOratio}
\eea
with $z=m_c^2/m_b^2$ and $f(z)=1-8z+8z^3-z^4-12z^2{\rm ln}z$.  

     The obtained branching ratio at the LO has non-negligible perturbative 
uncertainties, which are reduced by taking 
into account corrections at the next leading order (NLO).  
For the numerical evaluation, therefore, we incorporate NLO 
corrections for the matrix 
elements at $\mu=m_b$ \cite{ali} and the anomalous 
dimensions \cite{chetyrkin}.  
Our calculations follow formulae given in Ref. \cite{kagan},  
which also include QED corrections.  

     The decay width of $\BSgamma$ depends on the $U$-quark mass  
$m_U$ and the CKM matrix elements $\VT$, $\VU$, $\ZBS$.  
The value of $\ZBS$ determines the FCNC interactions at the 
tree level in Eq. (\ref{Zlagrangian}), which is constrained 
from non-observation of $B\rightarrow K\mu^+\mu^-$ \cite{pdg} as 
\beq
 |\ZBS| <8.1\times 10^{-4}.  
\eeq
The CKM matrix elements connecting light ordinary quarks, 
which are directly measured in experiments, have the same values 
as those in the SM.  
From the values of $V_{12}$, $V_{13}$, $V_{22}$, and $V_{23}$ \cite{pdg},  
we obtain a constraint 
\beq
0.03<|\VT+\VU-\ZBS|<0.05.  
\label{constraint}
\eeq
The mass $m_U$ should be heavier than the $t$-quark mass.  
In principle, the $U$-quark mass and the CKM matrix 
elements are not independent each other, their relations 
being determined by the mass matrices $M^u$ and $M^d$.  
However, these relations depend on many unknown factors 
for the mass matrices.  Furthermore, the values of 
$m_U$ and $V_{42}^*V_{43}$ are thoroughly unknown 
phenomenologically except for the above constraints.  
We therefore take them for independent parameters.  

\begin{figure}
\psfig{file=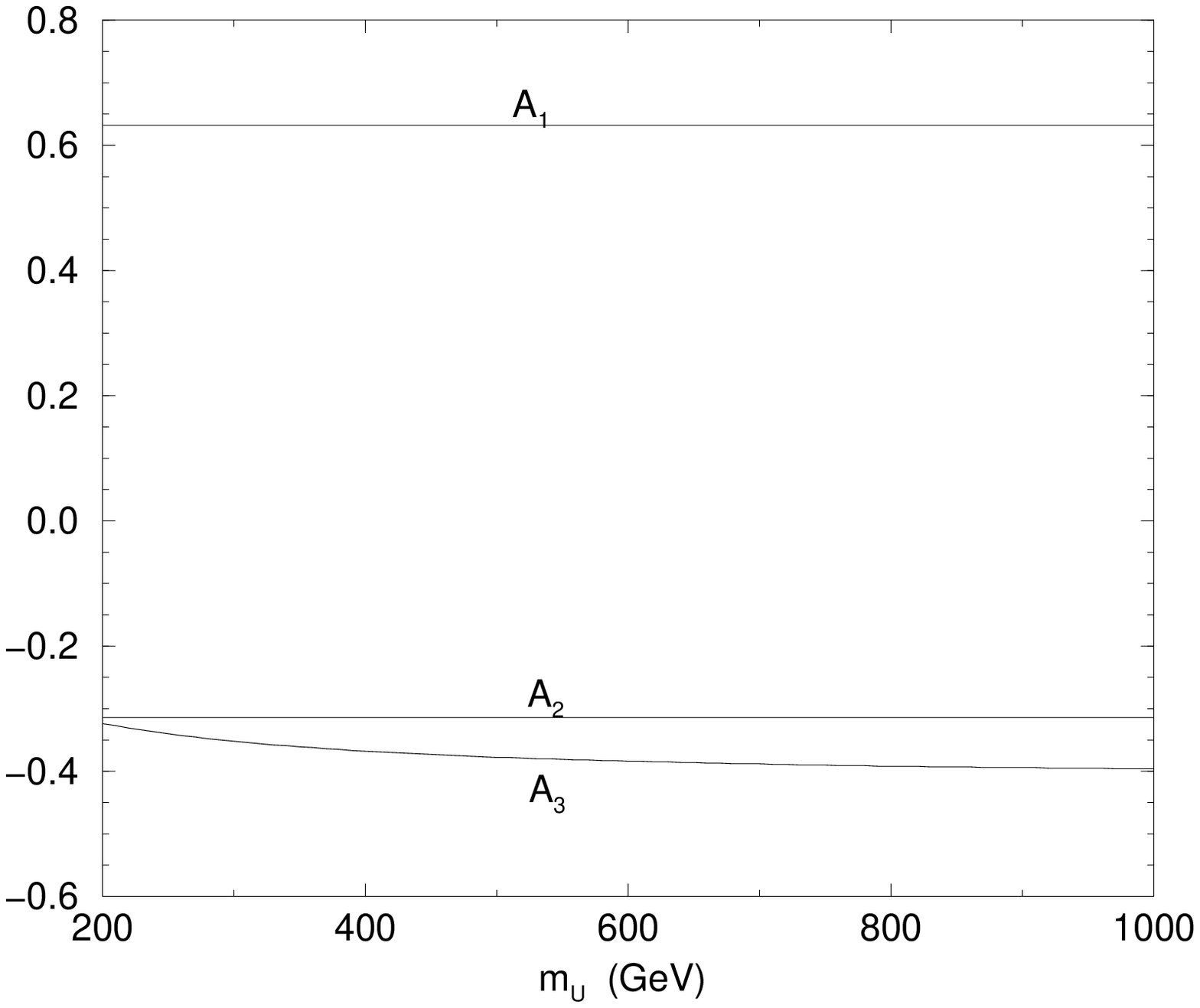,width=14cm,height=12cm,angle=0}
\caption{The values of $A_1$, $A_2$, and $A_3$ in Eq. (\ref{c7mod}).} 
\label{figmass}
\end{figure}

     The decay width is mainly determined by 
the Wilson coefficient $C_7(m_b)$ as seen from Eq. (\ref{LOratio}).  
Expressing explicitly the dependence on the CKM matrix elements, 
the coefficient $C_7(m_b)$ in Eq. (\ref{c7}) is written as 
\begin{equation}
 C_7(m_b)=A_1\ZBS+A_2\VT+A_3\VU,  
\label{c7mod}
\end{equation}
where $A_3$ is a function of $m_U$ while $A_1$ and $A_2$ are constants.  
We show the $m_U$ dependency of $A_3$ in Fig. \ref{figmass}, 
where $A_1$ and $A_2$ are also depicted.  
For $m_U\gsim 200$ GeV, the value of $A_3$ does not 
vary much with $m_U$ and is comparable with $A_2$.  
Unless $\VU$ is much smaller than $\VT$, 
the coefficient $C_7(m_b)$ can be predicted differently from the SM value.  
Although $A_1$ is larger than $A_2$ and $A_3$ in magnitude, 
the smallness of $\ZBS$ makes the term $A_1\ZBS$ less important.  

\begin{figure}
\psfig{file=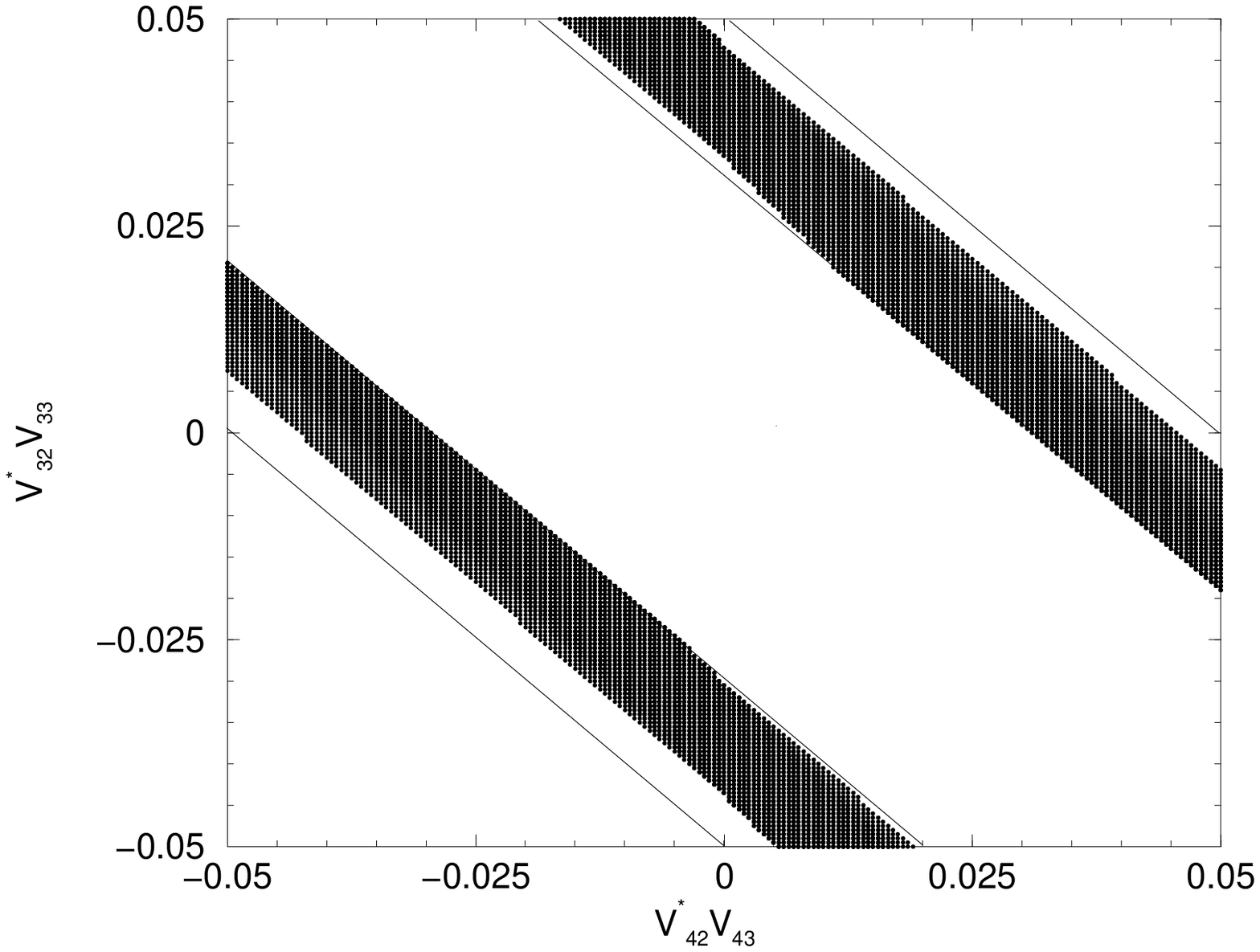,width=14cm,height=12cm,angle=0}
\caption{The allowed regions for $\VT$ and $\VU$.  $\ZBS=8.1\times 10^{-4}$.} 
\label{figckm}
\end{figure}

   In Fig. \ref{figckm} we show allowed regions for $\VT$ and $\VU$, 
assuming for simplicity that these values are real.  
The shaded regions are compatible with the experimental 
results of both Eq. (\ref{aleph}) for $\BSgamma$ and Eq. (\ref{constraint}) 
for the CKM matrix elements.   
The regions between the solid lines satisfy the latter.  
We have taken the $U$-quark mass for 200 GeV $<m_U<$ 1 TeV 
and $\ZBS$ for its maximal value $8.1\times 10^{-4}$.   
The branching ratio of $\BSgamma$ sizably constrains the 
CKM matrix elements of the VQM.  
The allowed regions are slightly altered for 
$\ZBS=-8.1\times 10^{-4}$.   

\begin{figure}
\psfig{file=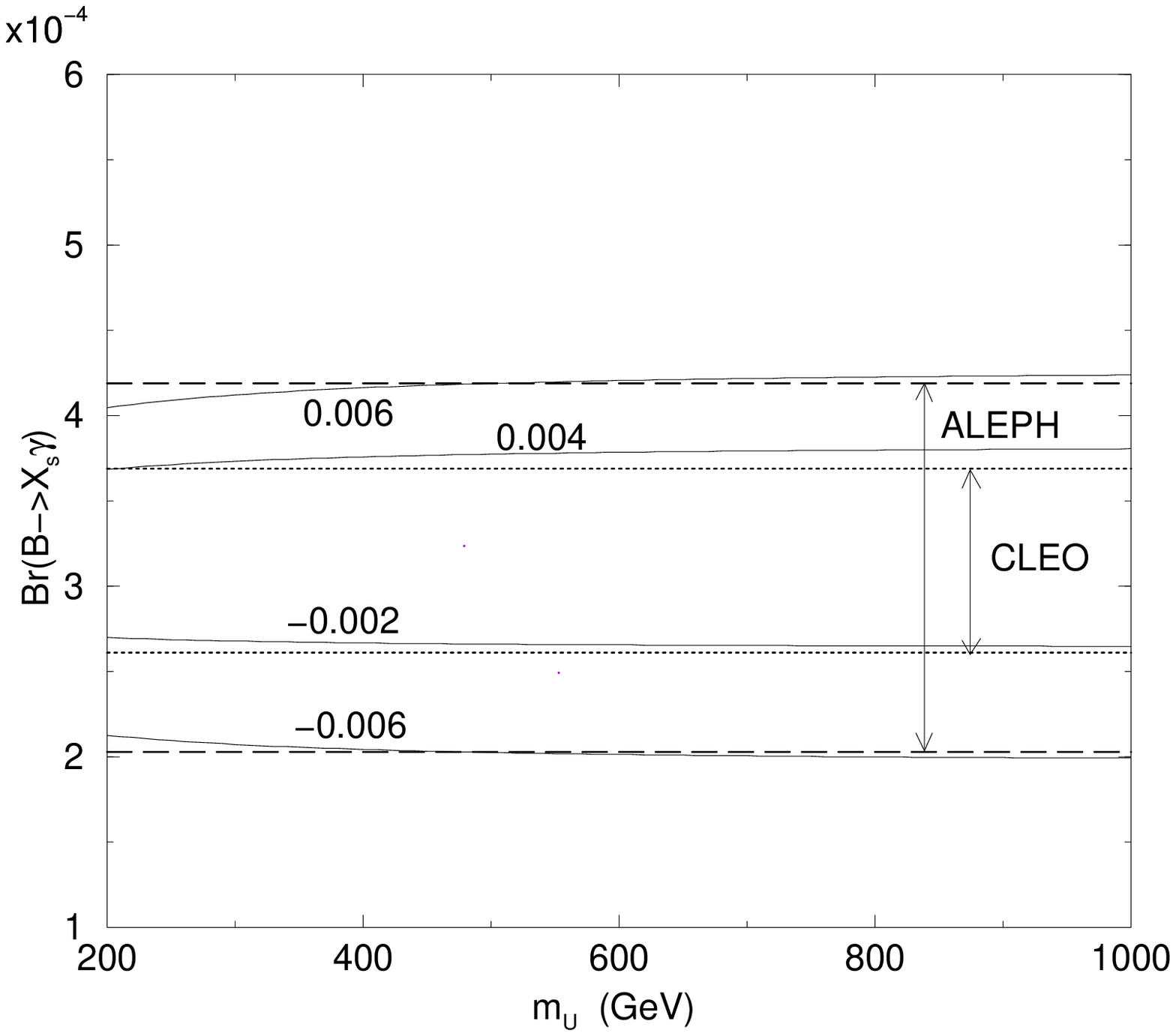,width=14cm,height=12cm,angle=0}
\caption{The branching ratio of $\BSgamma$.  
$\VU=-0.006, -0.002, 0.004, 0.006$, $\VT=0.04$, $\ZBS=8.1\times 10^{-4}$.} 
\label{figratio}
\end{figure}

     In Fig. \ref{figratio} the branching 
ratio of $\BSgamma$ is depicted as a function of $m_U$ for 
$\VU=-0.006, -0.002, 0.004, 0.006$.  
For definiteness, we put $\VT=0.04$ and $\ZBS=8.1\times 10^{-4}$.  
The experimental bounds Eqs. (\ref{cleo}) and (\ref{aleph}) are also shown.  
For $|\VU/\VT|\gsim 0.1$, the predicted value is non-trivially 
different from that of the SM.  
The branching ratio could have any value within the experimental bounds.  

     In summary, we have studied the effects of the VQM 
on the branching ratio for the radiative $B$-meson decay.
Among the possible new contributions, the $W$-mediated 
diagrams yield sizable effects.
From the experimental results for the branching ratio, 
the values of $\VT$ and $\VU$ are constrained.  
These constraints do not much depend on the mass of the 
extra quark $U$.  
The VQM could make the branching ratio of $\BSgamma$ different 
from the SM prediction.  
If precise measurements in the near future show a 
difference between the experimental value and the SM prediction, 
the VQM may become one candidate for physics beyond the SM.  

\smallskip 
 
     The authors thank G.C. Cho for discussions.       
This work is supported in part by the Grant-in-Aid for 
Scientific Research on Priority Areas (Physics of $CP$ Violation, 
No. 12014205) from the Ministry of Education, Science and
Culture, Japan.

\newpage 
 
\begin{table}
\begin{center}
\begin{tabular}{c c c c c c c c c}
\hline
 $i$   & 1 & 2 & 3 & 4 & 5 & 6 & 7 & 8 \\
\hline
   & & & & & & & & \\
 $a_i$  & $\frac{14}{23}$ & $\frac{16}{23}$ & $\frac{6}{23}$ & $-\frac{12}{23}$\

         & 0.4086 & $-0.4230$ & $-0.8994$ & 0.1456 \\
   & & & & & & & & \\
 $h_i$  & $\frac{626126}{272277}$ & $-\frac{56281}{51730}$ & $-\frac{3}{7}$
       & $-\frac{1}{14}$ & $-0.6494$ & $-0.0380$ & $-0.0186$ & $-0.0057$ \\
   & & & & & & & & \\
 $\bar h_i$  & $\frac{313063}{363036}$ & 0 & 0 & 0
       & $-0.9135$ & 0.0873 & $-0.0571$ & 0.0209   \\
   & & & & & & & & \\
\hline
\end{tabular}
\end{center}
\caption{The values of $h_i$, $\bar h_i$, and $a_i$ 
in Eqs. (\ref{c7}) and (\ref{c8}).}
\label{tabcoef}
\end{table}

\pagebreak

\pagebreak

\pagebreak

\end{document}